\pdfoutput=1
\documentclass[conference]{IEEEtran}
\usepackage{amsmath,amssymb}
\usepackage{graphicx}
\usepackage{booktabs}
\usepackage{multirow}
\usepackage{cite}
\usepackage{url}
\usepackage{array}
\usepackage{microtype}
\usepackage{siunitx}
\graphicspath{{wcnc_rewrite_assets/}}

\title{Utility-Aware Adaptive Downlink DM-RS Allocation\\
from Uplink CSI for Lightweight O-DU dApps}

\author{\IEEEauthorblockN{Seungwon Min}
\IEEEauthorblockA{Independent Researcher\\Republic of Korea}}

\usepackage{eso-pic}
\AddToShipoutPictureFG*{\AtPageUpperLeft{\put(38,-24){\parbox[t]{536pt}{\fontsize{7.5}{8.5}\selectfont
An extended follow-up, ``Receiver-Aware DM-RS Selection With Uplink CSI Statistics for O-RAN xApps,'' was submitted to IEEE Wireless Communications Letters on 31 October 2026.\\
This work has been submitted to the IEEE for possible publication. Copyright may be transferred without notice, after which this version may no longer be accessible.
}}}}

\begin{document}
\maketitle

\begin{abstract}
Adaptive demodulation reference signal (DM-RS) placement in 5G New Radio (NR) trades pilot overhead against channel-tracking robustness, making a single static density suboptimal across heterogeneous mobility and propagation regimes. This paper studies a lightweight per-UE controller intended for O-DU-side distributed application (dApp) execution. The controller consumes eleven deployable uplink-CSI statistics and selects among six standards-derived PDSCH DM-RS configurations. Rather than optimize oracle-class accuracy, we train an 11--128--64--32--6 multilayer perceptron with a utility-aware objective that preserves the throughput structure of all candidate actions. In a large-scale trajectory-disjoint evaluation covering 294 channel/speed/SNR conditions and 7,056 observations, the proposed U-SoftCE policy significantly improves throughput over a training-selected best fixed pattern by 1.30, 0.86, and 1.13 percentage points on in-distribution, CDL channel-OOD, and 300-km/h mobility-OOD splits, respectively; paired 95\% bootstrap intervals exclude zero in all three cases. An independent bit-accurate Sionna NR LDPC validation over 36 held-out conditions and 50 transport blocks per candidate retains a +1.21-point gain with a 95\% cluster-bootstrap interval of [0.27, 2.13]. A native C++ implementation uses 12,070 dense parameters and requires 5.59~$\mu$s mean inference time, while a FlexRIC-based emulated O-DU prototype demonstrates sub-10-ms closed-loop operation. These results support utility-aware objectives for compact adaptive PHY controllers while exposing the limits of exact class-accuracy optimization.
\end{abstract}


\section{Introduction}
PDSCH demodulation in NR relies on DM-RS observations to estimate the effective radio channel. Increasing DM-RS density improves tracking of time-selective channels, but each additional pilot consumes resource elements that could otherwise carry coded data. NR therefore provides multiple higher-layer-constrained DM-RS configurations rather than one universally optimal time density~\cite{3gpp38211}. The resulting engineering problem is not merely to identify the most frequent oracle class, but to select a pattern whose tracking benefit justifies its overhead under the current channel state.

Recent work has started to learn this decision. Ryu and Yang proposed channel-prediction-based reference signal allocation (CPRS), in which temporal CSI is processed by ViViT/CNN models to select standards-compliant downlink DM-RS allocations~\cite{ryu2026cprs}. Their results establish temporal channel information as a useful control signal for adaptive reference-signal placement. In parallel, the O-RAN dApp concept moves inference and lower-layer control closer to the O-DU/O-CU, complementing slower RIC loops~\cite{doro2022dapps,orangrg2026dapp}. FlexRIC provides a compact SDK and emulated E2-node environment for experimentally studying such programmable RAN control~\cite{schmidt2021flexric}.

A related AI-RAN perspective is provided by Yoo, Park, Park, and Kang~\cite{yoo2026airan}, who study test-time resource allocation for interacting AI modules in cell-free MIMO. Their formulation highlights that online AI decisions should adapt to the wireless channel and system state rather than use a fixed inference budget. Our problem is complementary: instead of allocating inference computation among multiple AI modules, we use locally available UL-CSI statistics to choose a PHY reference-signal action whose communication utility depends on the current channel.

A second issue is the learning objective. If the oracle pattern yields 760 useful bits/slot and another candidate yields 758 bits/slot, ordinary cross-entropy (CE) treats the latter as completely wrong. The same loss also treats a catastrophic 30-bit/slot decision as one wrong class. For PHY control, however, the cost of an error is determined by link utility rather than class identity.

This paper makes five contributions. First, we define a deployable 11-feature per-UE controller that excludes simulator-only labels. Second, we restrict the action space to six standards-derived Type-A PDSCH DM-RS configurations and select the best fixed baseline using training data only. Third, we evaluate utility-aware learning over 294 large-scale operating conditions with trajectory-level paired bootstrap inference. Fourth, we independently validate the learned policy with bit-accurate NR LDPC encoding/decoding. Fifth, we implement the compact model in C++ and evaluate a FlexRIC-based emulated O-DU closed loop, separating deployment-feasibility evidence from PHY performance claims.

\section{System Model and DM-RS Action Space}
\subsection{Link Utility}
For a received resource element,
\begin{equation}
 y_{t,k}=h_{t,k}x_{t,k}+w_{t,k},
\end{equation}
where $x_{t,k}$ is the transmitted symbol, $h_{t,k}$ the effective channel, and $w_{t,k}$ noise. Known DM-RS symbols provide observations for channel estimation and equalization. Let $\mathcal{P}$ denote the candidate set and let $\eta_p$ be useful throughput after the complete PHY processing chain for candidate $p$. The per-sample oracle is
\begin{equation}
 p^{\star}=\arg\max_{p\in\mathcal{P}}\eta_p,\qquad
 \eta^{\star}=\max_{p\in\mathcal{P}}\eta_p.
\end{equation}
The normalized candidate utility is
\begin{equation}
 u_p=\frac{\eta_p}{\eta^{\star}+\epsilon}\in[0,1].
\end{equation}
This target distinguishes a near-optimal action from a genuinely damaging one.

\subsection{Six Standards-Derived Candidates}
Table~\ref{tab:candidates} lists the six implemented actions. They are derived from the single- and double-symbol PDSCH DM-RS location tables in TS~38.211, Sec.~7.4.1.1~\cite{3gpp38211}, for the stated Type-A configuration with first DM-RS symbol $l_0=2$ and a full 14-symbol allocation. The study evaluates these configurations as a compact action set; it does not claim that arbitrary per-slot switching among all six is standardized for an unmodified commercial UE.

\begin{table}[t]
\caption{Six PDSCH DM-RS Candidate Configurations}
\label{tab:candidates}
\centering
\footnotesize
\begin{tabular}{@{}clll@{}}
\toprule
ID & Name & DM-RS symbols & Role \\
\midrule
0 & sparse\_1 & $\{2\}$ & minimum overhead \\
1 & medium\_2 & $\{2,11\}$ & moderate tracking \\
2 & dense\_3 & $\{2,7,11\}$ & denser tracking \\
3 & dense\_4 & $\{2,5,8,11\}$ & high time density \\
4 & double\_2 & $\{2,3\}$ & double-symbol, pos0 \\
5 & double\_4 & $\{2,3,10,11\}$ & double-symbol, pos1 \\
\bottomrule
\end{tabular}
\end{table}

\begin{figure}[t]
\centering
\includegraphics[width=\columnwidth]{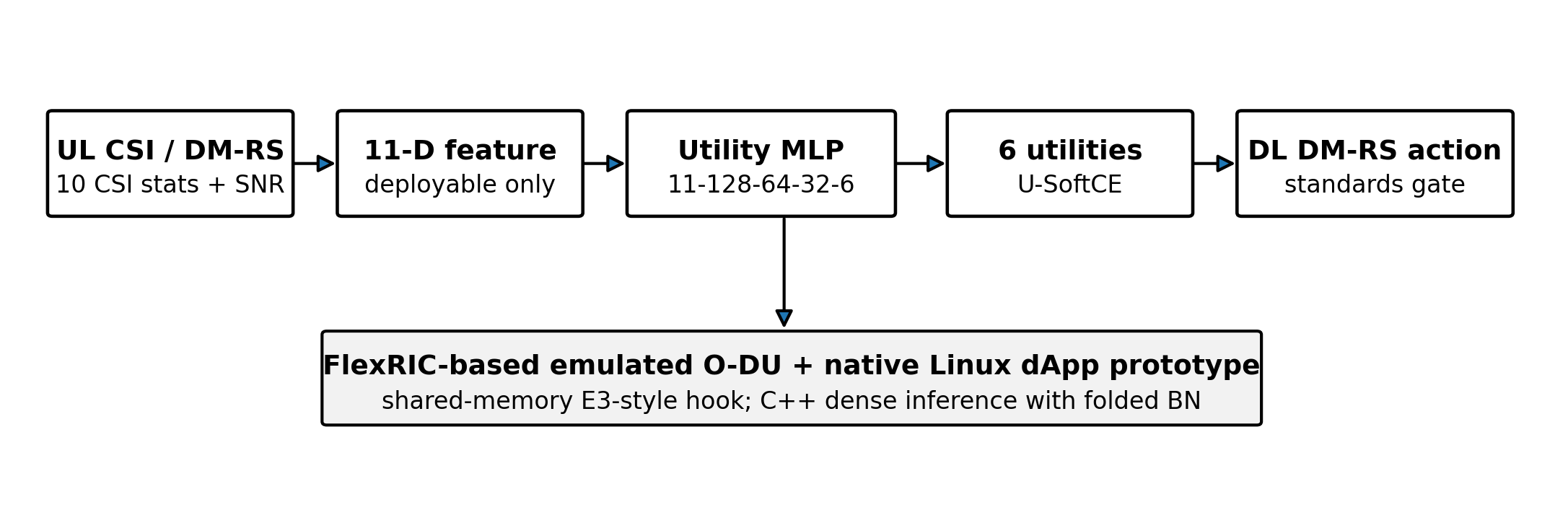}
\caption{Deployment-oriented controller and prototype path. Online inference uses only deployable UL-CSI features and an estimated SNR.}
\label{fig:architecture}
\end{figure}

\section{Utility-Aware Lightweight Controller}
\subsection{Deployable Inputs and Model}
The online feature vector contains eleven quantities obtainable from UL channel processing: mean UL power, an RMS-delay-spread proxy, frequency-correlation variance, a Doppler-spread proxy, spatial/subband power imbalance, phase-drift rate, subcarrier-variation coefficient, temporal autocorrelation, a spatial-condition-number proxy, mean phase deviation, and wideband effective SNR. Ground-truth speed, channel-model identity, LOS labels, and other simulator-only variables are excluded.

The controller is an MLP with dimensions $11\rightarrow128\rightarrow64\rightarrow32\rightarrow6$. Batch normalization follows each hidden affine layer. The dense weights and biases contain 12,070 parameters; three batch-normalization layers add 448 trainable affine parameters for 12,518 total during training. Batch normalization is folded into the preceding dense layers for the native C++ inference implementation.

\subsection{Utility-Soft Cross-Entropy}
We compare ordinary CE, a physics-based heuristic, direct utility regression (U-MSE), and U-SoftCE. U-MSE predicts $\hat{\mathbf{u}}\in\mathbb{R}^{6}$ and minimizes
\begin{equation}
 \mathcal{L}_{\mathrm{MSE}}=\frac{1}{|\mathcal{P}|}\sum_{p\in\mathcal{P}}(\hat{u}_p-u_p)^2.
\end{equation}
For U-SoftCE, normalized utilities are converted into soft targets
\begin{equation}
 q_p=\frac{u_p^{\gamma}}{\sum_j u_j^{\gamma}+\epsilon},\qquad \gamma=4,
\end{equation}
and the objective is
\begin{equation}
 \mathcal{L}_{\mathrm{U\text{-}SoftCE}}
 =\mathcal{L}_{\mathrm{MSE}}+\alpha\left[-\sum_p q_p\log\big(\mathrm{softmax}(\hat{\mathbf{u}})_p\big)\right],
\end{equation}
with $\alpha=0.1$. The selected action is $\arg\max_p\hat{u}_p$. Numerical and gradient unit tests were used to verify the implementation of this objective.

Models are trained with Adam (learning rate $3\times10^{-4}$, weight decay $10^{-4}$), batch size 64, and three independent random seeds. The best fixed policy, dense\_4, is selected on training data and then frozen for all test comparisons.

\section{Evaluation Methodology}
\subsection{Large-Scale Trajectory Evaluation}
The large-scale study contains 294 channel/speed/SNR operating conditions and 7,056 observations. The ID split contains 2,100/1,050/1,050 train/validation/test observations. Channel OOD holds out CDL-A/C conditions (2,016 observations), and mobility OOD evaluates 300~km/h (840 observations). The main metric is achieved throughput normalized by the per-sample oracle. Statistical uncertainty is estimated with 10,000 trajectory-level paired bootstrap resamples. Because the paired estimator gives equal weight to held-out trajectories, paired gains are reported directly rather than inferred by subtracting separately aggregated percentages.

The physics baseline is a coherence-time/Jakes-style rule. CE-MLP uses the same feature vector and architecture family but optimizes hard oracle labels. U-MSE and U-SoftCE share the candidate-utility targets, isolating whether the soft classification term provides additional benefit.

\subsection{Independent Bit-Accurate NR LDPC Validation}
The independent validation uses Sionna 1.2.2~\cite{sionna2022} with TensorFlow 2.21 and NR transport-block encoding/decoding. The evaluation integrates Sionna TBEncoder/TBDecoder with eight BP iterations and uses the same learned three-seed U-SoftCE ensemble; no random action bias is added at evaluation time. Thirty-six independent TDL-C/CDL-A, mobility, and SNR conditions are evaluated at 30, 120, and 300~km/h. Each of the six candidates is simulated with 50 LDPC transport blocks per condition. The policy is compared with the fixed dense\_4 action using a condition-level cluster bootstrap.

\subsection{Emulated O-DU Prototype and Network Stress Test}
A FlexRIC-based emulated O-DU E2 agent is paired with a native Linux dApp prototype. The fast path uses a shared-memory hook between the scheduler-side process and the controller. A Python closed-loop benchmark evaluates 1,000 slots, while a separately optimized C++ inference kernel measures the dense network alone. This distinction is retained in the reported latency numbers.

For a secondary system-level stress experiment, PHY-derived rate and BLER are mapped to Linux traffic control. HTB enforces the PHY rate, while a child \texttt{netem} qdisc injects loss and delay. TCP/UDP runs are repeated five times per scheme. This experiment intentionally omits HARQ/RLC recovery and is therefore treated as a network-emulation stress test, not a direct estimate of NR user-plane goodput.

\section{Results}
\subsection{Large-Scale Utility Performance}
\begin{figure*}[t]
\centering
\includegraphics[width=0.82\textwidth]{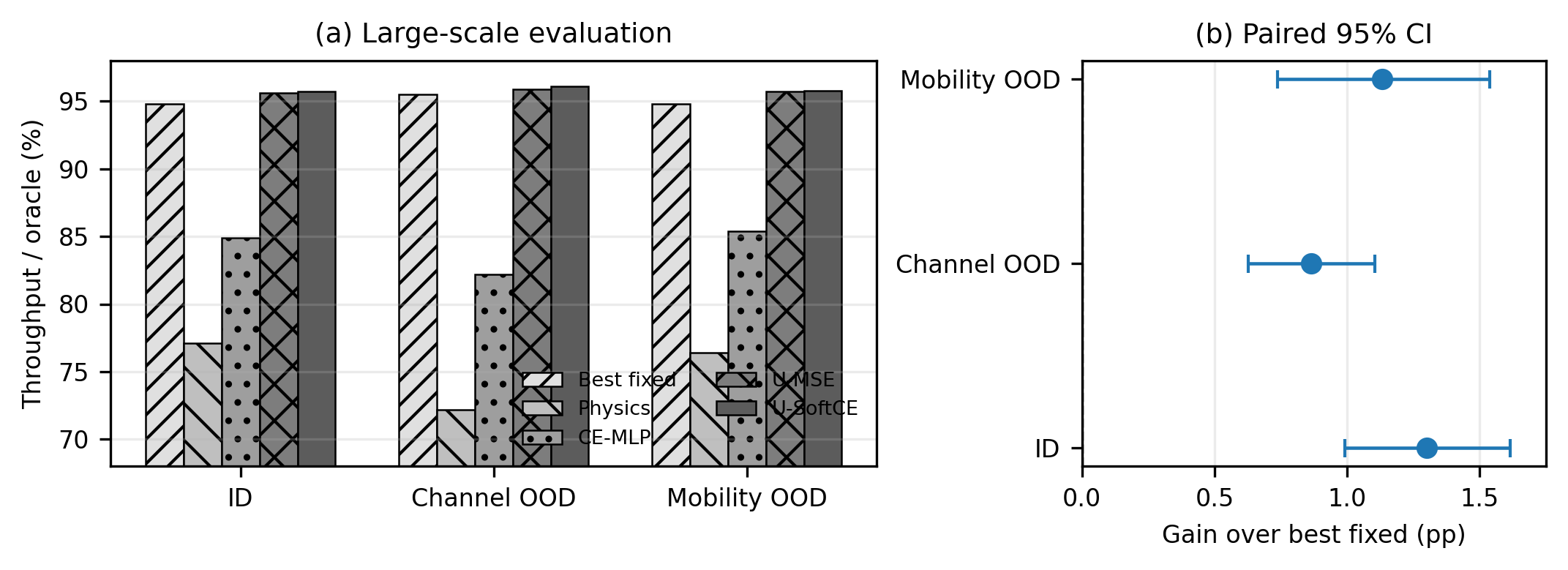}
\caption{Large-scale evaluation. (a) Normalized throughput across held-out regimes. (b) Trajectory-paired U-SoftCE gain over the training-selected best fixed pattern; all 95\% confidence intervals exclude zero.}
\label{fig:mainresults}
\end{figure*}

Across the three splits, U-SoftCE reaches 95.7\%, 96.1\%, and 95.8\% of oracle throughput, respectively, while the best fixed pattern averages 95.0\% across the study. More importantly, the trajectory-paired gain over the training-selected fixed baseline is positive and statistically significant on every split: +1.300~pp [0.990, 1.614] for ID, +0.863~pp [0.626, 1.103] for channel OOD, and +1.133~pp [0.738, 1.538] for 300-km/h mobility OOD. Figure~\ref{fig:mainresults} shows both the absolute normalized throughput and the paired confidence intervals.

\begin{table}[t]
\caption{Large-Scale Normalized Throughput (\% of Oracle)}
\label{tab:main}
\centering
\footnotesize
\begin{tabular}{@{}lrrrr@{}}
\toprule
Split & Fixed & CE & U-MSE & U-SoftCE \\
\midrule
ID & 94.8 & 84.9 & 95.6 & \textbf{95.7} \\
Channel OOD & 95.5 & 82.2 & 95.9 & \textbf{96.1} \\
Mobility OOD & 94.8 & 85.4 & 95.7 & \textbf{95.8} \\
Average & 95.0 & 84.2 & 95.7 & \textbf{95.9} \\
\bottomrule
\end{tabular}
\end{table}

\subsection{Utility-Aware Learning, Not Exact Class Accuracy}
The accuracy-throughput mismatch is pronounced. CE-MLP obtains top-1 accuracies of 45.0\%, 44.3\%, and 42.4\% on ID, channel-OOD, and mobility-OOD, while U-SoftCE obtains only 28.0\%, 34.0\%, and 31.0\%. Nevertheless, U-SoftCE achieves 10.8--13.9 percentage points higher normalized throughput than CE across the same splits. Figure~\ref{fig:validation}(a) illustrates this inversion. The result supports the use of communication utility rather than exact oracle-class recovery as the learning target.

The paired U-SoftCE versus U-MSE ablation further narrows the claim. U-SoftCE improves over U-MSE on ID by +0.372~pp with a 95\% CI of [0.098, 0.689], but the differences on channel OOD (-0.001~pp [-0.131, 0.140]) and mobility OOD (+0.043~pp [-0.137, 0.223]) are not significant. Thus, the robust conclusion is that utility-aware learning is substantially better aligned with throughput than hard-label CE; the specific U-SoftCE formulation provides an additional verified advantage only on ID in the present data.

\subsection{Bit-Accurate NR LDPC Validation}
\begin{figure}[t]
\centering
\includegraphics[width=0.90\columnwidth]{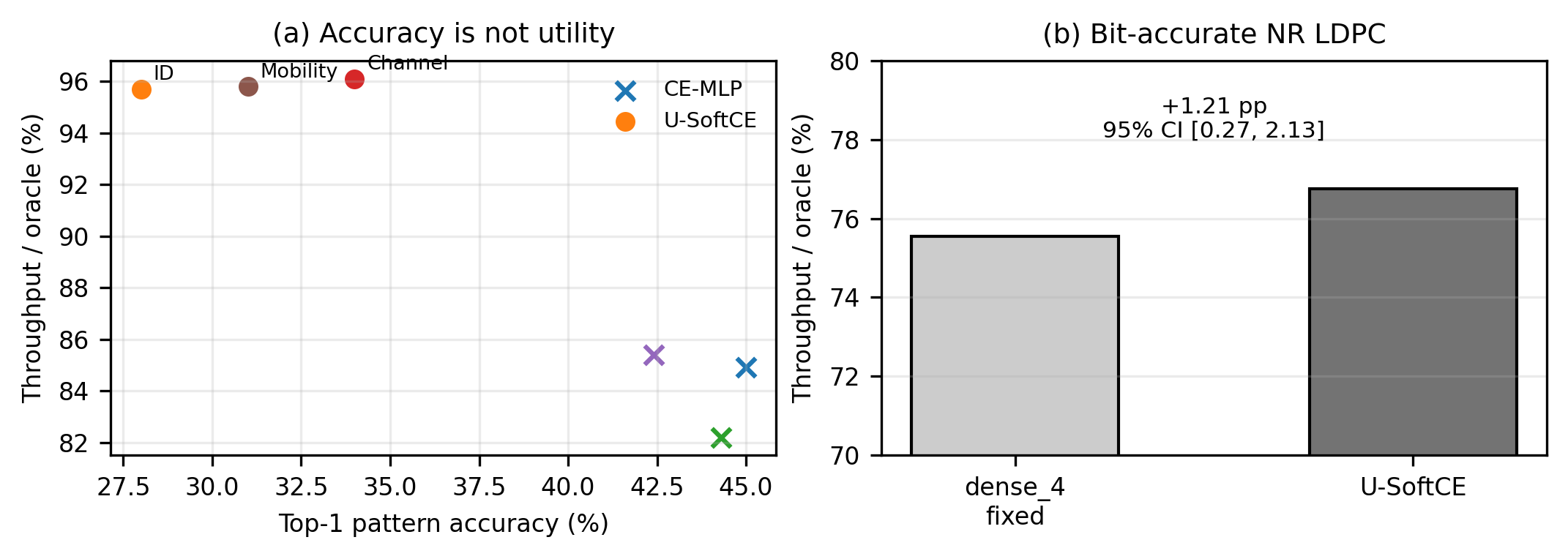}
\caption{(a) Top-1 pattern accuracy versus communication utility. (b) Independent bit-accurate Sionna NR LDPC validation.}
\label{fig:validation}
\end{figure}

The independent bit-accurate experiment confirms that the large-scale result is not solely an artifact of the fast link abstraction. U-SoftCE achieves 76.76\% of oracle throughput versus 75.54\% for dense\_4. The condition-level paired gain is +1.21~pp with a 95\% cluster-bootstrap interval of [0.27, 2.13], which excludes zero. At 300~km/h, the gain is +1.72~pp. Figure~\ref{fig:validation}(b) summarizes the overall comparison. The wider absolute confidence intervals relative to the large-scale study reflect the smaller, more computationally expensive bit-accurate grid.

\subsection{Prototype Latency and PHY-Informed Network Emulation}
The Python shared-memory closed-loop prototype has 0.522-ms mean RTT, 0.813-ms p95, 1.220-ms p99, and a 1.818-ms maximum over 1,000 observed slots. It therefore satisfies a 10-ms dApp control budget but does not meet a strict 1-ms p99 target. The optimized C++ dense-inference kernel is substantially smaller in scope and requires 5.59~$\mu$s on average, with a 31.91~$\mu$s observed maximum over 500 runs.

The corrected HTB+\texttt{netem} experiment enforces physically plausible rates; for example, a 4.30-Mbit/s PHY limit produces 4.225~Mbit/s in a validation TCP run. Under the 300-km/h stress setting, five repeated 8-s runs give mean UDP delivered goodputs of 5.918, 5.970, and 6.302~Mbit/s for fixed, U-SoftCE, and oracle, respectively. TCP variance is large (0.188$\pm$0.097, 0.249$\pm$0.087, and 0.224$\pm$0.095~Mbit/s), so we do not claim a transport-layer ranking from this experiment. The purpose is to verify rate enforcement and expose the learned PHY policy to a reproducible network-stack stress path.

\section{Discussion}
\subsection{Relation to CPRS and AI-RAN}
CPRS~\cite{ryu2026cprs} uses temporal CSI and a substantially richer ViViT/CNN channel-prediction pipeline to optimize standards-compliant reference-signal allocation. Our objective is complementary: we study a compact per-UE controller that consumes low-dimensional deployable statistics and directly optimizes candidate-level communication utility. A direct numerical comparison with CPRS is not claimed because its original ray-tracing dataset and implementation are not reproduced here.

Yoo et al.~\cite{yoo2026airan} formulate channel-aware test-time inference-resource allocation for cell-free MIMO AI-RANs. Their work motivates the broader principle that online AI decisions should be conditioned on wireless and system state. The present study applies that principle at a different layer: the online decision is a per-UE DM-RS action, and the optimized resource is PHY pilot overhead rather than inference compute.

\subsection{Limitations}
Four limitations bound the claims. First, the six actions are standards-derived under the stated PDSCH Type-A configuration; this study does not demonstrate arbitrary per-slot reconfiguration signaling to an unmodified commercial UE. Second, the scalable evaluation uses a computational link abstraction, while the independent Sionna validation covers 36 conditions and 50 transport blocks per candidate; broader bit-accurate sweeps would further reduce Monte-Carlo uncertainty. Third, the current O-DU path is a FlexRIC-based emulated E2 agent with a native Linux dApp hook rather than an integrated OAI/srsRAN production O-DU. Fourth, the HTB/\texttt{netem} stress test injects PHY-derived loss without HARQ/RLC, so its TCP/UDP values are illustrative network-emulation outputs rather than direct NR user-plane predictions.

A further methodological limitation is that U-SoftCE and U-MSE are statistically indistinguishable on both OOD splits. Hence, the present evidence supports utility-aware learning as the primary contribution; superiority of the soft-classification term itself is established only on ID.

\section{Conclusion}
This paper revisited adaptive downlink DM-RS selection as a communication-utility optimization problem rather than an oracle-class classification problem. A 12.5k-parameter controller using only deployable UL-CSI statistics selects among six standards-derived PDSCH DM-RS configurations. In a large-scale trajectory-disjoint study, U-SoftCE significantly outperforms the training-selected best fixed pattern on ID, CDL channel-OOD, and 300-km/h mobility-OOD splits. A separate bit-accurate NR LDPC validation retains a +1.21-percentage-point gain with a 95\% cluster-bootstrap interval excluding zero. The accuracy-throughput mismatch shows why exact pattern labels are a poor surrogate for PHY utility, while the U-MSE ablation indicates that the broader utility-aware formulation is more robustly supported than any single loss variant. Finally, a native C++ implementation and FlexRIC-based emulated O-DU prototype demonstrate that the controller is computationally compatible with lower-layer dApp execution. Future work should integrate the controller into a real O-DU scheduler/PHY path and validate dynamic DM-RS signaling with full HARQ/RLC and commercial-UE constraints.

\section*{Acknowledgment}
Generative-AI disclosure: OpenAI ChatGPT was used to assist with manuscript drafting, language refinement, LaTeX formatting, and figure-layout preparation. The author independently defined the study, generated and verified the simulation and prototype results, checked the technical claims and references, and assumes full responsibility for the manuscript.


\begin{thebibliography}{9}
\bibitem{3gpp38211}
3GPP, ``NR; Physical channels and modulation,'' 3GPP TS 38.211, Release 18.

\bibitem{3gpp38901}
3GPP, ``Study on channel model for frequencies from 0.5 to 100 GHz,'' 3GPP TR 38.901, Release 18.

\bibitem{ryu2026cprs}
S. Ryu and H. J. Yang, ``Standards-Compliant DM-RS Allocation via Temporal Channel Prediction for Massive MIMO Systems,'' \emph{IEEE Trans. Veh. Technol.}, vol. 75, no. 7, pp. 15171--15175, Jul. 2026, doi: 10.1109/TVT.2026.3654149.

\bibitem{doro2022dapps}
S. D'Oro, M. Polese, L. Bonati, H. Cheng, and T. Melodia, ``dApps: Distributed Applications for Real-Time Inference and Control in O-RAN,'' \emph{IEEE Commun. Mag.}, vol. 60, no. 11, pp. 52--58, Nov. 2022, doi: 10.1109/MCOM.002.2200079.

\bibitem{orangrg2026dapp}
O-RAN ALLIANCE Next Generation Research Group, ``dApp Architecture and Interfaces,'' Research Report, 2026.

\bibitem{yoo2026airan}
S. Yoo, S. Park, S.-H. Park, and J. Kang, ``Test-Time Scalable AI-RAN: Inference Time Allocation for Cell-Free MIMO,'' arXiv:2608.03614, 2026.

\bibitem{sionna2022}
J. Hoydis, S. Cammerer, F. Ait Aoudia, A. Vem, N. Binder, G. Marcus, and A. Keller, ``Sionna: An Open-Source Library for Next-Generation Physical Layer Research,'' arXiv:2203.11854, 2022.

\bibitem{schmidt2021flexric}
R. Schmidt, M. Irazabal, and N. Nikaein, ``FlexRIC: An SDK for Next-Generation SD-RANs,'' in \emph{Proc. ACM CoNEXT}, 2021, pp. 411--425, doi: 10.1145/3485983.3494870.
\end{thebibliography}
\end{document}